\documentclass[11pt]{article}

\usepackage{graphicx}
\usepackage{rotating}

\newcommand{\BABARPubYear}    {06}

\newcommand{\BABARConfNumber} {041}
\newcommand{\SLACPubNumber} {12009}

\input pubboard/babarsym
\RequirePackage{xspace}

\newcommand{\calP}{\ensuremath{{\cal P}}}

\newcommand{\pvec}{{\bf p}}

\providecommand{\skz}{\mbox{$S$}}
\providecommand{\ckz}{\mbox{$C$}}
\def\deltaS{\ensuremath{{\rm \Delta}S}\xspace}

\newcommand{\DE}{\ensuremath{\Delta E}}
\newcommand{\mb}{\ensuremath{m_{\rm ES}}}

\newcommand{\xf}{\ensuremath{{\cal F}}}

\newcommand{\thetaT}{\ensuremath{\theta_{\rm T}}}
\newcommand{\costhr}{\ensuremath{\cos\thetaT}}

\providecommand{\dt}{\deltat}
\newcommand{\ttag}{\ensuremath{t_{\rm tag}}}
\newcommand{\bflav}{\ensuremath{B_{\rm flav}}}

\newcommand\etal{{\it et al.}}
\newcommand{\half}{\ensuremath{{1\over2}}}

\newcommand{\msp}{\ensuremath{\phantom{-}}}

\newcommand{\bfig}{\begin{figure}[htbpc!]}
\newcommand{\efig}{\end{figure}}
\newcommand\bef{\begin{figure}}
\newcommand\edf{\end{figure}}
\newcommand\dbline{\noalign{\vskip 0.10truecm\hrule}\noalign{\vskip 2pt}\noalign{\hrule\vskip 0.10truecm}}

\newcommand\sgline{\noalign{\vskip 0.10truecm\hrule\vskip 0.10truecm}}
\newcommand\beq{\begin{equation}}
\newcommand\eeq{\end{equation}}
\newcommand\bear{\begin{array}}
\newcommand\enar{\end{array}}
\newcommand\beqa{\begin{eqnarray}}
\newcommand\eeqa{\end{eqnarray}}
\newcommand\ben{\begin{enumerate}}
\newcommand\een{\end{enumerate}}

\newcommand{\UfourS}{\ensuremath{\Upsilon(4S)}}

\newcommand{\KSpm}{\ensuremath{K^0_{{\scriptscriptstyle S}+-}}\xspace}
\newcommand{\KSzz}{\ensuremath{K^0_{{\scriptscriptstyle S}00}}\xspace}

\newcommand{\etagg}{\ensuremath{\eta_{\gaga}}}
\newcommand{\etappp}{\ensuremath{\eta_{3\pi}}}

\newcommand{\etapepp}{\ensuremath{\etapr_{\eta\pi\pi}}}

\newcommand{\etaprg}{\ensuremath{\etapr_{\rho\gamma}}}

\newcommand{\etapeppgg}{\ensuremath{\etapr_{\eta(\gamma\gamma)\pi\pi}}}
\newcommand{\etapeppppp}{\ensuremath{\etapr_{\eta(3\pi)\pi\pi}}}

   \newcommand{\rhoz}{\ensuremath{\rho^0}}

   \newcommand{\fetaprgKp}{\ensuremath{\etapr_{\rho\gamma} K^+}}

\newcommand{\fetapKz}{\ensuremath{\etapr K^0}}
\newcommand{\fetapKs}{\ensuremath{\etapr\KS}}
\newcommand{\fetapKl}{\ensuremath{\etapr\KL}}
\newcommand{\etapKz}{\ensuremath{\Bz\ra\fetapKz}}
\newcommand{\etapKs}{\ensuremath{\Bz\ra\fetapKs}}
\newcommand{\etapKl}{\ensuremath{\Bz\ra\fetapKl}}

   \newcommand{\fetaprgKz}{\ensuremath{\etapr_{\rho\gamma} K^0}}

\providecommand{\tcp}{\mbox{$t_{CP}$}}
\providecommand{\sigdt}{\ensuremath{\sigma_{\deltat}}}

\newcommand{\fetapreppggkz}{\ensuremath{\etapr_{\eta(\gamma\gamma)\pi\pi} K^{0} }}
\newcommand{\fetapreppggk}{\ensuremath{\etapr_{\eta(\gamma\gamma)\pi\pi} K^{+} }}
\newcommand{\fetapreppthrpikz}{\ensuremath{\etapr_{\eta(3\pi)\pi\pi}K^{0}}}
\newcommand{\fetapreppthrpik}{\ensuremath{\etapr_{\eta(3\pi)\pi\pi} K^{+}}}

\newcommand{\fetapreppkzl}{\ensuremath{\etapr_{\etagg\pi\pi} \KL}}

\newcommand{\SetapKz}{\ensuremath{0.55\pm0.11\pm0.02}}
\newcommand{\CetapKz}{\ensuremath{-0.15\pm0.07\pm0.03}}

\long\def\inst#1{\par\nobreak\kern 4pt\nobreak
    {\it #1}\par\vskip 10pt plus 3pt minus 3pt}

\begin{document}
{\pagestyle{empty}

\begin{flushright}
SLAC-PUB-\SLACPubNumber \\
\babar-CONF-\BABARPubYear/\BABARConfNumber \\
\end{flushright}

\par\vskip 5cm

\begin{center}
{\Large \bf\boldmath 
Time-dependent $CP$-violation parameters in \etapKz\ decay
}
\end{center}
\bigskip

\begin{center}
\large The \babar\ Collaboration\\
\mbox{ }\\
\today
\end{center}
\bigskip \bigskip

\begin{center}
\large \bf Abstract
\end{center}
We present measurements of time-dependent \CP-violation asymmetries for
the decays \etapKz.  The data sample corresponds to 347 million \BB\
pairs produced by \epem\ annihilation at the \UfourS\ resonance in the
PEP-II collider, and collected with the \babar\ detector.  The
preliminary results are $\skz = \SetapKz$, and $\ckz = \CetapKz$, where
the first error quoted is statistical, the second systematic.

\vfill

\begin{center}

Submitted to the 33$^{\rm rd}$ International Conference on High-Energy Physics, ICHEP 06,\\
26 July---2 August 2006, Moscow, Russia.

\end{center}

\vspace{1.0cm}
\begin{center}
{\em Stanford Linear Accelerator Center, Stanford University, 
Stanford, CA 94309} \\ \vspace{0.1cm}\hrule\vspace{0.1cm}
Work supported in part by Department of Energy contract DE-AC03-76SF00515.
\end{center}

\newpage
} 

\begin{center}
\small

The \babar\ Collaboration,
\bigskip

%
{B.~Aubert,}
{R.~Barate,}
{M.~Bona,}
{D.~Boutigny,}
{F.~Couderc,}
{Y.~Karyotakis,}
{J.~P.~Lees,}
{V.~Poireau,}
{V.~Tisserand,}
{A.~Zghiche}
\inst{Laboratoire de Physique des Particules, IN2P3/CNRS et Universit\'e de Savoie,
 F-74941 Annecy-Le-Vieux, France }
% ...

\end{center}\newpage

\section{INTRODUCTION}
\label{sec:Introduction}

Measurements of time-dependent \CP\ asymmetries in $B^0$ meson decays
through a dominant Cabibbo-Kobayashi-Maskawa (CKM) favored $b \rightarrow c
\bar{c} s$ amplitude \cite{babar} have provided a crucial test of the
mechanism of \CP\ violation in the Standard Model (SM) \cite{SM}.
For such decays the interference between this amplitude and \BzBzb\
mixing is dominated by the single phase $\beta = \arg{(-V_{cd} V^*_{cb}/
V_{td} V^*_{tb})}$ of the CKM mixing matrix.
Decays of $B^0$ mesons to charmless hadronic final states such as
$\etapr K^0$ proceed mostly via a single loop (penguin)
amplitude with the same weak phase as the $b \to c \bar{c} s$ transition
\cite{Penguin}, but
CKM-suppressed 
amplitudes and multiple particles in the loop introduce additional weak
phases whose contribution may not be negligible
\cite{Gross,Gronau,Gronau2,BN,london}.

For the decay \etapKz, these additional contributions are expected to be
small within the SM, so the time-dependent asymmetry measurement for
this decay provides an approximate measurement of \stwob.  Theoretical
bounds for the small deviation \deltaS\ between the time-dependent
\CP-violation parameter $S$ measured in this decay and in the
charmonium-\Kz decays have been calculated with an SU(3) analysis \cite
{Gross,Gronau} from measurements of \Bz\ decays to pairs of neutral
light pseudoscalar mesons \cite{Isosca,PRD04}.  The most stringent of
these is given by Eq. 19 in \cite{Gronau}, which assumes negligible
contributions from exchange and penguin annihilation, and has a
theoretical uncertainty less than $\sim$0.03.  With newer measurements
\cite{Isosca06}\ we obtain an improved bound $\deltaS<0.08$
\cite{bbKzKstz}.  QCD factorization calculations 
conclude that \deltaS\ is even smaller \cite{BN}.  A significantly
larger \deltaS\ could arise from non-SM amplitudes \cite{london}.

The time-dependent \CP-violation asymmetry in the decay \etapKz\ has
been measured previously by the \babar ~\cite{Previous} and
Belle~\cite{BELLE,belles2b} experiments.  In this paper we update our previous
measurements with an improved analysis and a data sample 1.5 times
larger.

\section{THE \babar\ DETECTOR AND DATASET}
\label{sec:babar}

The data were collected with the \babar\ detector~\cite{BABARNIM} at the
PEP-II asymmetric-energy \epem\ collider \cite{pep}.  An integrated
luminosity of 316~fb$^{-1}$, corresponding to 347 million \BB\ pairs,
was recorded at the $\Upsilon (4S)$ resonance (center-of-mass energy
$\sqrt{s}=10.58\ \gev$).  

Charged particles from \epem\ interactions are detected, and their
momenta measured, by a combination of five layers of double-sided
silicon microstrip detectors and a 40-layer drift chamber,
both operating in the 1.5~T magnetic field of a superconducting
solenoid. Photons and electrons are identified with a CsI(Tl)
electromagnetic calorimeter (EMC).  Further charged particle
identification (PID) is provided by the average energy loss ($dE/dx$) in
the tracking devices and by an internally reflecting ring imaging
Cherenkov detector (DIRC) covering the central region.  The instrumented 
flux return (IFR) of the magnet allows discrimination of
muons from pions.

\section{ANALYSIS METHOD}
\label{sec:Analysis}

\subsection{\boldmath Time evolution of a \BzBzb\ pair}
\label{sec:TD}

From a candidate \BB\ pair we reconstruct a \Bz decaying into the
\CP\ eigenstate $f=\fetapKs$ or $f=\fetapKl$ ($B_{CP}$).  From the
remaining particles in the event we also reconstruct the 
vertex of the other $B$ meson ($B_{\rm tag}$) and identify its flavor.
The difference $\deltat \equiv \tcp - \ttag$ of the proper decay times
$\tcp$ and $\ttag$ of the signal and tag $B$ mesons, respectively, is
obtained from the measured distance between the $B_{CP}$ and $B_{\rm
tag}$ decay vertices and from the boost ($\beta \gamma =0.56$) of the
\epem system.
The \deltat\ distribution is given by:
\begin{equation}
  F(\dt) =
        \frac{e^{-\left|\deltat\right|/\tau}}{4\tau} [1 \mp\Delta w \pm
   (1-2w)\left(S\sin(\deltamd\deltat) -
   C\cos(\deltamd\deltat)\right)].  \label{eq:FCPdef}
\end{equation}
The upper (lower) sign denotes a decay accompanied by a \Bz (\Bzb) tag,
$\tau$ is the mean $\Bz$ lifetime, $\deltamd$ is the mixing frequency,
and the mistag parameters $w$ and $\Delta w$ are the average and
difference, respectively, of the probabilities that a true $\Bz$\ is
incorrectly tagged as a $\Bzb$\ or vice versa.  The tagging algorithm
\cite{s2b} has six mutually exclusive tagging categories based on
quantities such as the
sign of charge of a lepton, kaon, or soft pion from \Dstar,
grouped according to their
response purities.  The measured analyzing power, defined as
efficiency times $(1-2w)^2$ summed over all categories, is $( 30.4\pm
0.3)\%$, as determined from a large sample of $B$-decays to fully
reconstructed flavor eigenstates (\bflav).  The parameter $C$ measures
direct \CP violation.  If $C=0$, then $S=-\eta \stwob+\deltaS$, where
$\eta$ is the \CP eigenvalue of the final state ($-1$ for \fetapKs,
$+1$ for \fetapKl).

\subsection{Event selection}
\label{sec:EvtSel}

We establish the event selection criteria with the aid of a detailed
Monte Carlo (MC) simulation of the \B\ production and decay sequences,
and of the detector response \cite{geant}.  These criteria are designed
to retain signal events with high efficiency.  Applied to the data, they
result in a sample much larger than the expected signal, but with well
characterized backgrounds. We extract the signal yields from this sample
with a maximum likelihood (ML) fit.

\begin{table}[btp]
\begin{center}
\caption{
Selection requirements on the invariant masses of resonances and the
laboratory energies of photons from their decay.}
\label{tab:rescuts}
\begin{tabular}{lcc}
\dbline
State		& Invariant mass (MeV)			& $E(\gamma)$ (MeV)\\
\sgline						
\piz\ {\rm (from \etappp)}& $120 < m(\gamma\gamma) < 150$	& $>30$	\\
\piz\ {\rm (from \KSzz)}  & $120 < m(\gamma\gamma) < 155$	& $>30$	\\
\etagg		& $490 < m(\gamma\gamma) < 600$		& $>50$	\\
\etappp		& $520 < m(\pip\pim\piz) < 570$		& ---	\\
\etapepp	& $945 < m(\pip\pim\eta) <970$		& ---	\\
\etaprg		& $930 < m(\pip\pim\gamma) <980$	& $>100$	\\
\rhoz		& $470 < m(\pip\pim) <980$		& ---	\\
\KSpm		& $486 < m(\pip\pim) <510$		& ---	\\
\KSzz		& $468 < m(\piz\piz) <528$		& ---	\\
\dbline
\end{tabular}
\vspace{-5mm}
\end{center}
\end{table}

The \B-daughter candidates are reconstructed through their decays
$\piz\ra\gaga$, $\eta\ra\gaga$ (\etagg), $\eta\ra\pip\pim\piz$
(\etappp), $\etapr\ra\etagg\pip\pim$ (\etapeppgg), 
$\etapr\ra\etappp\pip\pim$ (\etapeppppp),
$\etapr\ra\rhoz\gamma$
(\etaprg), where $\rhoz\ra\pip\pim$, $\KS\ra\pip\pim$ (\KSpm) or
$\piz\piz$ (\KSzz).  Table \ref{tab:rescuts}\ lists the
requirements on the invariant mass of these particles' final states.
Secondary charged pions in \etapr\ and $\eta$ candidates are rejected if
classified as protons, kaons, or electrons by their DIRC, $dE/dx$, and
EMC PID signatures.  We require \KS\ candidates to have a flight length
with significance $>$3$\sigma$. 
Signal \KL\ candidates are reconstructed from clusters of energy
deposited in the EMC or from hits in the IFR not associated with any
charged track in the event.  From the cluster centroid and the \Bz
decay vertex we determine the direction (but not the magnitude) of the
$\KL$ momentum $\pvec_{\KL}$.

For decays with a \KS\ we reconstruct the \B-meson candidate by
combining the four-momenta of the \KS\ and \etapr\ and imposing a vertex
constraint.  Since
the natural widths of the $\eta$, \etapr, and \piz\ are much smaller
than the resolution, we also constrain their masses to world-average values
\cite{PDG2006}\ in the fit of the \B\ candidate.
From the kinematics of \UfourS\ decay we determine the
energy-substituted mass $\mes \equiv \sqrt{(\half s +
\pvec_0\cdot\pvec_B)^2/E_0^2 - \pvec_B^2}$ and the energy difference
$\DE \equiv E_B^*-\half\sqrt{s}$, where $(E_0,\pvec_0)$ and
$(E_B,\pvec_B)$ are four-momenta of the \UfourS\ and the $B$ candidate,
respectively, and the asterisk denotes the \UfourS\ rest frame.  The
resolution in \mes\ is $3.0\ \mev$ and in \DE\ is $20-50$ MeV,
depending on the decay mode.  We require $5.25<\mes<5.29\ \gev$
and $|\DE|<0.2$ GeV ($-0.01<\DE<0.08$ GeV for \etapKl).

For a \etapKl\ candidate we obtain \DE\ and $p_{\KL}$ from a fit with
$B^0$ and \KL\ masses constrained to their accepted
values~\cite{PDG2006}.  To make a match with the measured \KL\ direction
we construct the missing momentum $\pvec_{\rm
miss}$ from $\pvec_0$ and all charged tracks and neutral clusters other
than the \KL.  We then project $\pvec_{\rm miss}$ onto $\pvec_{\KL}$, and
require the component perpendicular to the beam line, 
$\pvec_{{\rm miss}\perp}^{\rm proj}$, to satisfy
$p_{{\rm miss}\perp}^{\rm proj}-p_{\KL\perp} > -0.5\ \gevc$.  This value
reflects the resolution, and was chosen to minimize the yield
uncertainty in the presence of continuum background.

For all \etapKz\ candidates we require for \dt\
and its error \sigdt, $|\dt|<20$ ps and $\sigdt<2.5$ ps.

\subsection{Background rejection}
\label{sec:Bkg}

Backgrounds arise primarily from random combinations of particles in
continuum $\epem\ra\qqbar$ events ($q=u,d,s,c$).  We reduce these with
requirements on the angle \thetaT\ between the thrust axis of the $B$
candidate in the \UfourS\ frame and that of the rest of the charged
tracks and neutral calorimeter clusters in the event.  The distribution
is sharply peaked near $|\costhr|=1$ for \qqbar\ jet pairs, and nearly
uniform for $B$-meson decays.  The requirement, which optimizes the
expected signal yield relative to its background-dominated statistical
error, is $|\costhr|<0.9$ ($|\costhr|<0.8$ for \etapKl).

In the ML fit we discriminate further against \qqbar\ background with a
Fisher discriminant \xf\ that combines several variables which
characterize the production dynamics and energy flow in the event
\cite{PRD04}.  It provides 
about one standard deviation of separation between \B\ decay events and
combinatorial background.

For the \etaprg\ decays we require $|\cos\theta^{\rho}_{\rm dec}|< 0.9$
to exclude the most asymmetric decays where soft-particle backgrounds
concentrate and the acceptance changes rapidly.  Here
$\theta_{\rm dec}^\rho$ is the angle between the momenta of the \rhoz\
daughter \pim\ and the \etapr, measured in the \rhoz\ rest frame.

For \etapKl\ candidates we require that the cosine of the
polar angle of the total missing momentum in the laboratory
system be less than 0.95, to reject very forward \qqbar\ jets.
The purity of the \KL\ candidates reconstructed in the EMC is further
improved by a requirement on the output of a neural network (NN) that
takes cluster-shape variables as its inputs. The NN was trained on MC
signal events and data events in the sideband \mbox{$0.04<\DE<0.08$
\gev}.  We checked the performance of the NN with \KL\ candidates in the
larger $\Bz \to J/\psi K^0_L$ sample. 

The average number of candidates found per 
selected event is in the range 1.08 to 1.32, depending on the final
state. 
We choose the candidate with the smallest value of a $\chi^2$ 
constructed from the deviations from expected values of one or more of
the daughter resonance masses, or with the best vertex probability for the
$B$, depending on the decay channel.  In  \etapKl\ if several $B$
candidates have the same vertex  
probability, we chose the candidate with the \KL\ reconstructed from, in
order, EMC and IFR, EMC only, or IFR only.
From the simulation we find that
this algorithm selects the correct-combination candidate in about
two thirds of the events containing multiple candidates, and that it
induces negligible bias.

\subsection{Maximum likelihood fit}
\label{sec:MLfit}

We obtain the common \CP-violation parameters and yields for each
channel from a maximum likelihood fit with the input observables \DE,
\mes, \xf, and \deltat.  The selected sample sizes are given in the
first column of Table~\ref{tab:Results}.  Besides the signal events they
contain \qqbar\ (dominant) and \bbbar\ with $b\ra c$ combinatorial
background, and a fraction that we estimate from the simulation to be
less than 1.1\% of cross feed from other charmless \BB\ modes.  The
charmless events (henceforth refered to as \BB) have ultimate final
states different from the signal, but 
similar kinematics, and exhibit broad peaks in the signal regions of
some observables.  We account for these with a separate component in the
probability density function (PDF).  For each component $j$ (signal,
\qqbar\ combinatorial background, or \BB\ background) and tagging
category $c$, we define a total probability density function for
event $i$ as
\begin{equation}
{\cal P}_{j,c}^i \equiv {\cal P}_j ( \mes^i ) \cdot {\cal
P}_j ( \DE^i ) \cdot { \cal P}_j( \xf^i ) \cdot 
{ \cal  P}_j (\deltat^i, \sigma_{\deltat}^i;c)\,,
\end{equation}
except for \etapKl\ for which ${\cal P}_j ( \mes^i )$ is omitted.
The factored form of the PDF is a good approximation, particularly for the
combinatorial \qqbar\ component, since correlations among observables
measured in the data (in which \qqbar\ dominates) are small.  Distortions
of the fit results caused 
by this approximation are measured in simulation and
included in the bias corrections and systematic errors discussed below.

With $Y_j$ defined to be the yield of events of component $j$,
and $f_{j,c}$ the fraction of events of component $j$ for each category $c$,
we write the extended likelihood function for all events belonging to
category $c$ as 
\begin{equation}
{\cal L}_c = \exp{\Big(-\sum_{j} Y_jf_{j,c}\Big)}
  \prod_i^{N_c} (Y_{\rm sig}f_{{\rm sig},c}{\cal P}_{{\rm sig},c}^{i}
  +Y_{q\bar{q}} f_{q\bar{q},c}{\cal P}_{q\bar{q}}^{i}
  +Y_{B\bar{B}}f_{B\bar{B},c}{\cal P}_{B\bar{B}}^{i}),
\end{equation}
where $N_c$ is the number of events of category $c$ in the sample. We
found that the \BB\ background component is needed only for the channels
with \etaprg.
We fix both $f_{{\rm sig},c}$ and 
$f_{B\bar{B},c}$ to $f_{\bflav,c}$, the values measured with the large
\bflav\ sample  \cite{Resol}. 
The total likelihood function ${\cal L}_d$ for decay mode $d$ is given as the
product over the six tagging categories.  Finally, when combining
decay modes we form the grand likelihood ${\cal L}=\prod{\cal L}_d$. 

The PDF $\calP_{\rm sig}(\dt,\, \sigdt, c)$ is given by
$F(\dt)$ (Eq.\ \ref{eq:FCPdef}) with tag category ($c$) dependent mistag
parameters convolved with the
signal resolution function (a sum of three Gaussians) determined from the
\bflav\ sample. We determine the remaining PDFs for the signal and \BB\
background components from fits to MC data, for which the resolutions in
\DE\ and \mes\ are calibrated with large control samples of $B$ decays
to charmed final states of similar topology (e.g.\ $B\ra
D(K\pi\pi)\pi$).  For the combinatorial background the PDFs are
determined in the fits to the data.  However we first deduce the
functional form from a fit of each component alone to a sideband in
(\mes,\,\DE), so that we can validate the fit before applying it to data
containing the signal.

These PDF forms are: the sum of two Gaussians for ${\cal P}_{\rm
sig}(\mes)$ and ${\cal P}_{\rm sig}(\DE)$; the sum of three Gaussians
for ${\cal P}_{\qqbar}(\dt; c)$; a conjunction of two Gaussian segments
below and above the peak with different widths for ${\cal P}_j(\xf)$ (a
small ``tail" Gaussian is added for ${\cal P}_{\qqbar}(\xf)$); a linear
dependence for ${\cal P}_{\qqbar}(\DE)$; and for ${\cal
P}_{\qqbar}(\mes)$ the function
$x\sqrt{1-x^2}\exp{\left[-\xi(1-x^2)\right]}$, with
$x\equiv2\mes/\sqrt{s}$.  These are discussed in more detail in
\cite{PRD04}.

We allow the parameters most important for the determination of the
combinatorial background PDFs to vary in the fit.  Thus for the six
channels listed in Table \ref{tab:Results} we perform a single fit with
109 free parameters: $-\eta S$, $C$, signal yields (6),
$\etapr_{\rho\gamma} K^0$ \BB\ background yields (2), continuum
background yields (6) and fractions (30), background \dt,
\mes, \DE, \xf\ PDF parameters (63). The parameters $\tau$ and 
$\deltamd$ are fixed to world-average values \cite{PDG2006}.  The 
symbol $S$ refers to $S_{\fetapKs}$, and inclusion of the \CP
eigenvalue $\eta$ of the final state accounts for the expected
difference in sign with respect to \fetapKl. 

We test and calibrate the fitting procedure by applying it to
ensembles of simulated \qqbar\ experiments drawn from the PDF into which
we have embedded the expected number of signal and \BB\ background
events randomly extracted from the fully simulated MC samples.  We find
negligible bias for $C$.  For $S$ we find and apply multiplicative
correction factors
for bias from dilution due to \BB\ background, equal to 1.02 in 
the final states $\fetaprgKz_{\pi^+\pi^-}$ and \fetapreppkzl, and 1.05 in
$\fetaprgKz_{\pi^{0}\pi^{0}}$. 

\subsection{Fit Results}
\label{sec:fitResults}

\begin{table}[!hb]
\caption{Results with statistical errors for the $\Bz\to\etapr K^0$
time-dependent fits.}
\label{tab:Results}
\begin{center}
\vspace*{-0.3cm}
\begin{tabular}{lcccc}
\hline\hline
Mode                          &Events to fit  &Signal yield  & $-\eta S$       &         $C$      \\
\hline
$\fetapreppggkz_{\pi^+\pi^-}$	  &612   &$206\pm16$  &$\msp0.60\pm0.24$ &$-0.26\pm0.14$ \\
$\fetaprgKz_{\pi^+\pi^-}$	  &10905 &$503\pm28$  &$\msp0.50\pm0.15$ &$-0.26\pm0.11$ \\
$\fetapreppthrpikz_{\pi^+\pi^-}$  &164   &$63\pm8$    &$\msp0.85\pm0.38$ &$\msp0.24\pm0.26$ \\
$\fetapreppggkz_{\pi^{0}\pi^{0}}$ &446   &$50\pm9$    &$\msp0.77\pm0.44$ &$-0.25\pm0.36$ \\
$\fetaprgKz_{\pi^{0}\pi^{0}}$     &12559 &$114\pm23$  &$\msp0.42\pm0.47$ &$\msp0.30\pm0.30$ \\
\hline
$\fetapKs$                        &      &            &$\msp0.57\pm0.11$ &$-0.18\pm0.08$ \\
\hline
$\fetapKl$                        &3389  &$168\pm21$  &$\msp0.39\pm0.30$ &$\msp0.20\pm0.23$ \\
\hline
$\fetapKz$                        &      &            &$\msp0.55\pm0.11$ &$-0.15\pm0.07$ \\
\hline\hline
\end{tabular}
\end{center}
\vspace*{-0.3cm}
\end{table}

\begin{figure}[!htb]
\begin{center}
 \includegraphics[angle=0,width=0.9\textwidth]{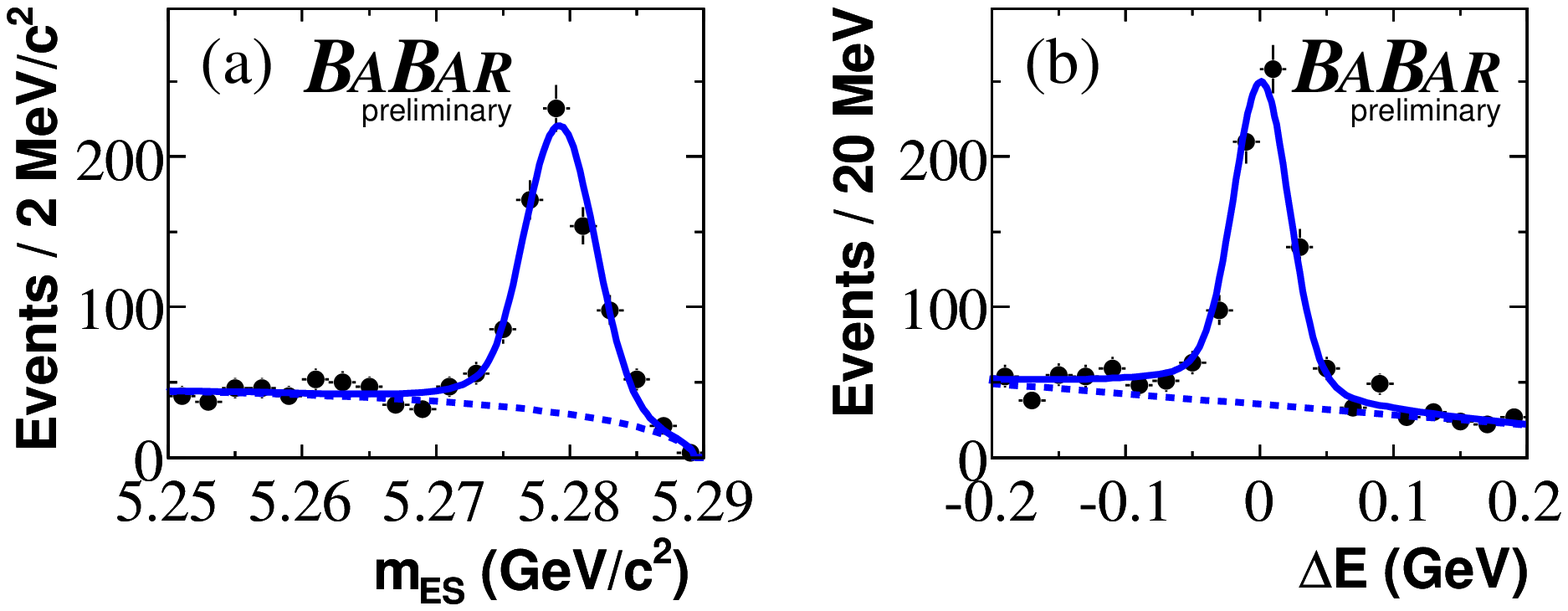}\\
\vspace{-0.7cm}
\end{center}
 \caption{\label{fig:projMbDE}
Distributions projected onto (a) \mb\ and (b) \DE\ for
\etapKs\ candidates. }
\end{figure}

\begin{figure}[!htb]
\begin{center}
 \includegraphics[angle=0,width=0.55\textwidth]{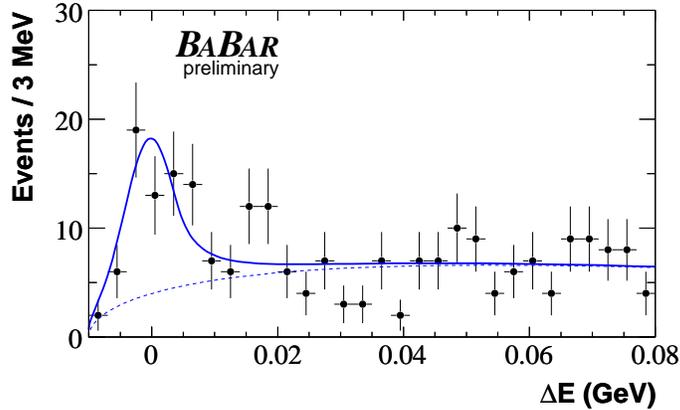}
\vspace{-0.7cm}
\end{center}
 \caption{\label{fig:projDE_KL}
Distribution projected onto \DE\ for \etapKl\ candidates.  Points with error 
bars represent the data, the solid line the fit function, and the dashed
line its background component. }
\end{figure}

\begin{figure}[!htb]
  \begin{center}
   \includegraphics[width=0.49\textwidth]{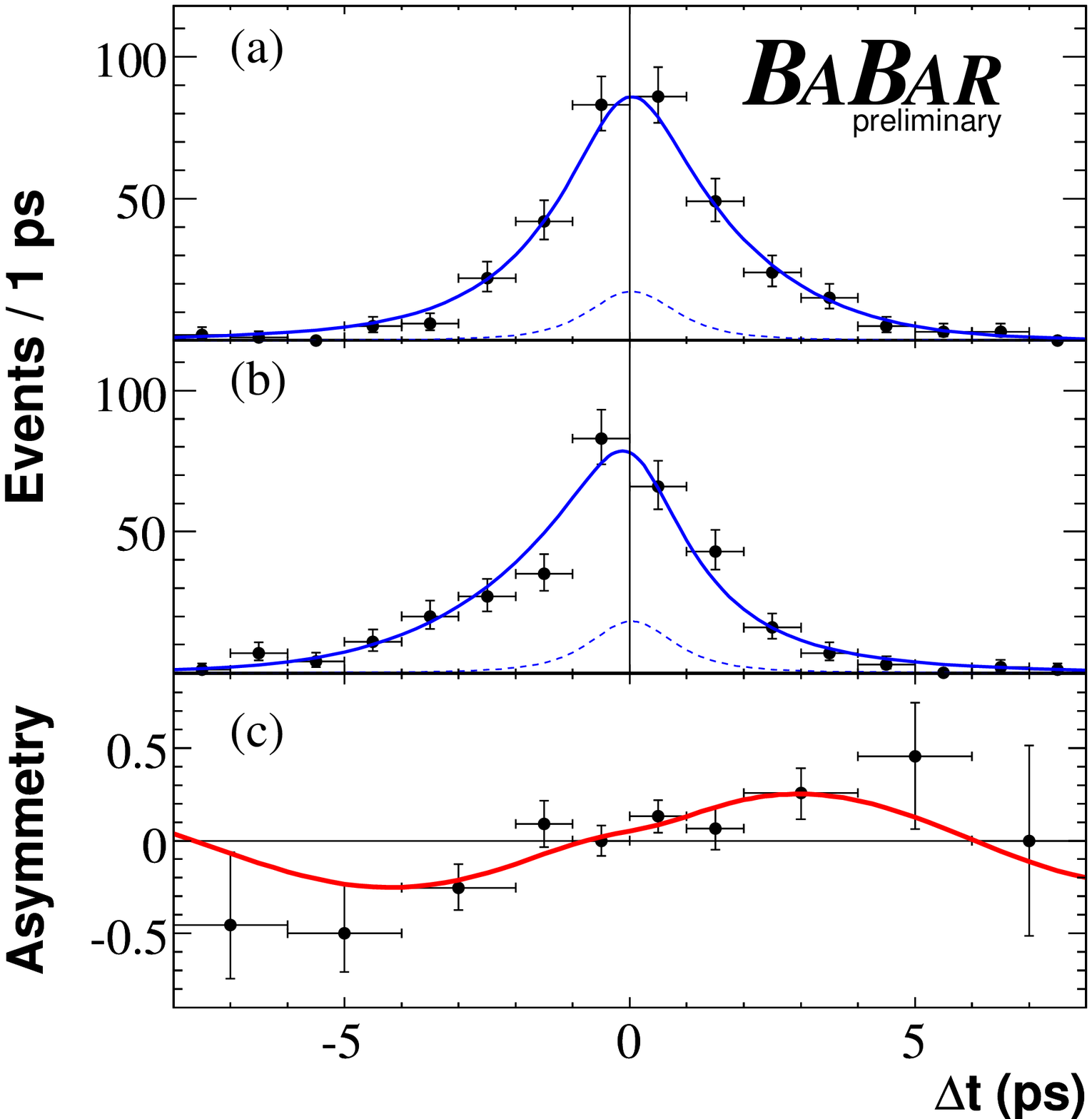}
   \includegraphics[width=0.49\textwidth]{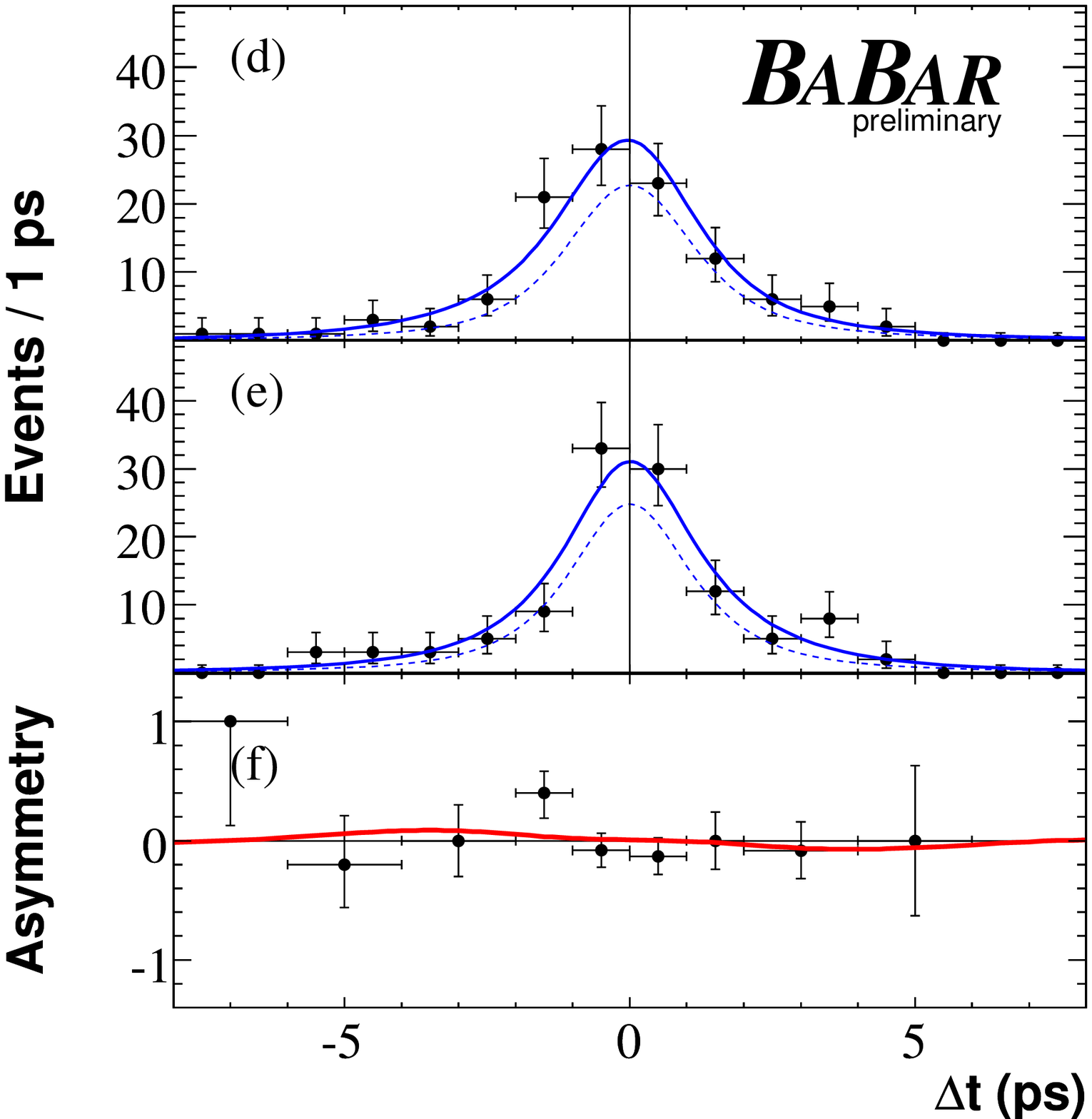}
\end{center}
  \vspace*{-0.7cm}
 \caption{Projections onto $\Delta t$ for (a-c) \etapKs\ and (d-f) 
\etapKl\ of the data (points with  error bars), fit function (solid 
line), and background function (dashed line), for (a, d) \Bz\ and (b, e) 
\Bzb\ tagged events, and (c, f) the asymmetry between \Bz\ and \Bzb\ 
tags.}
  \label{fig:DeltaTProj}
\end{figure}

Results from the fit for the signal yields and the \CP parameters $S$
and $C$ are presented in Table \ref{tab:Results}.
In Fig.\ \ref{fig:projMbDE}\ we show for \etapKs\ the projections onto
\mb\ and \DE\ for 
a subset of the data for which the signal likelihood
(computed without the variable plotted) exceeds a mode-dependent
threshold that optimizes the sensitivity; the corresponding
distribution in \DE\ for \etapKl\ is given in Fig.\ \ref{fig:projDE_KL}.
Fig.~\ref{fig:DeltaTProj} gives the $\Delta t$
projections and asymmetry of the combined modes for 
events selected as for Figs.~\ref{fig:projMbDE} and \ref{fig:projDE_KL}.
We measure a correlation of $3.0\%$ between $S$ and $C$ in the fit.

We perform numerous crosschecks of our fitter: time-dependent 
fits for $B^+$
decays to the charged final states $\fetapreppggk$, $\fetaprgKp$, 
and $\fetapreppthrpik$; fits removing one fit variable
at a time; fits without \BB\ PDFs; fits with multiple \BB\ components; 
fits allowing for non-zero \CP information in \BB events; fits with $C = 0$ 
and others. In all cases, we find results consistent with expectation. 
The value
$S_{\fetapKs}=0.57\pm0.11$ is larger than our previous measurement
$S_{\fetapKs}=0.30\pm0.14$ \cite{Previous} as a result of the larger
data sample and events added or removed as a result
of changes in the reconstruction and selection. For events common to the
two datasets we find close agreement of the values of $S$ and $C$.

\subsection{Systematic studies}
\label{sec:Systematics}

We find
systematic uncertainties from several sources (in decreasing order of
magnitude): variation of the signal
PDF shape parameters within their errors, modeling of the signal \dt\
distribution, use of \dt\ signal parameters
from the \bflav\ sample, interference between the CKM-suppressed
$\bar{b}\to\bar{u} c\bar{d}$ amplitude and the favored $b\to c\bar{u}d$
amplitude for some tag-side $B$ decays
\cite{dcsd}, \BB\ background, SVT alignment, and position and
size of the beam spot. The \bflav\ sample is used  to determine the errors
associated with the signal \dt\ resolutions, tagging efficiencies, and
mistag rates.  We take the uncertainties in $\tau_B$ and \deltamd\ from 
the published measurements \cite{PDG2006}. 
Summing all systematic errors in quadrature, we obtain 0.02 
for $S$ and 0.03 for $C$.

\section{RESULTS AND DISCUSSION}
\label{sec:Summary}

In conclusion, we have used samples of about 940 \etapKs\ and 170 \etapKl\
events to measure the time-dependent \CP violation parameters in
\etapKz\  $\skz = \SetapKz$ and $\ckz = \CetapKz$.  Our result for
\skz\ is consistent with the world average of those measured in $\Bz\ra
J/\psi\KS$ \cite{s2b,belles2b}, and inconsistent with zero (\CP
conservation) by 4.9 standard deviations.  Our result for the
direct-\CP
parameter \ckz\ is 1.8 standard deviations from zero.  The results
are preliminary.

\section{ACKNOWLEDGMENTS}
\label{sec:Acknowledgments}

We are grateful for the 
extraordinary contributions of our \pep2\ colleagues in
achieving the excellent luminosity and machine conditions
that have made this work possible.
The success of this project also relies critically on the 
expertise and dedication of the computing organizations that 
support \babar.
The collaborating institutions wish to thank 
SLAC for its support and the kind hospitality extended to them. 
This work is supported by the
US Department of Energy
and National Science Foundation, the
Natural Sciences and Engineering Research Council (Canada),
Institute of High Energy Physics (China), the
Commissariat \`a l'Energie Atomique and
Institut National de Physique Nucl\'eaire et de Physique des Particules
(France), the
Bundesministerium f\"ur Bildung und Forschung and
Deutsche Forschungsgemeinschaft
(Germany), the
Istituto Nazionale di Fisica Nucleare (Italy),
the Foundation for Fundamental Research on Matter (The Netherlands),
the Research Council of Norway, the
Ministry of Science and Technology of the Russian Federation, 
Ministerio de Educaci\'on y Ciencia (Spain), and the
Particle Physics and Astronomy Research Council (United Kingdom). 
Individuals have received support from 
the Marie-Curie IEF program (European Union) and
the A. P. Sloan Foundation.

\end{document}